\documentclass[journal]{IEEEtran}
\usepackage{color,graphicx}
\usepackage[applemac]{inputenc}
\usepackage{epsfig}
\usepackage{latexsym}
\usepackage{amsfonts}
\usepackage{amssymb}
\usepackage{amsmath}
\usepackage{amsbsy}
\usepackage{color,graphicx}
\usepackage{multirow}
\usepackage{float}
\usepackage[english]{babel}
\usepackage{url}
\usepackage{subfigure}
\usepackage{ulem}
\usepackage[square, comma, sort&compress]{natbib}

\graphicspath{{./}{./fig/}}
\usepackage{enumerate}

\newcommand{\argmax}[2]{\smash{\mathop{{\rm argmax}}\limits_{#1}}\, #2 } 
\newcommand{\argmin}[2]{\smash{\mathop{{\rm argmin}}\limits_{#1}}\, #2 } 

\DeclareTextFontCommand{\emph}{\itshape}

\ifCLASSINFOpdf
\else

\fi

\begin{document}

\title{
%
Statistics of the MLE and Approximate Upper and Lower Bounds -- Part 2: Threshold Computation and Optimal Signal Design
}


\author{\authorblockN{Achraf~Mallat,~\IEEEmembership{Member,~IEEE,}
				Sinan~Gezici,~\IEEEmembership{Senior Member,~IEEE,}
				Davide~Dardari,~\IEEEmembership{Senior Member,~IEEE,}
        and~Luc~Vandendorpe,~\IEEEmembership{Fellow,~IEEE}}
\thanks{Achraf Mallat and Luc Vandendorpe are with the ICTEAM Institute, Universit\'e Catholique de Louvain, Belgium. Email:
\{Achraf.Mallat, Luc.Vandendorpe\}@uclouvain.be.}
\thanks{Sinan Gezici is with the Department of Electrical and Electronics Engineering, Bilkent University, Ankara 06800, Turkey. Email:
gezici@ee.bilkent.edu.tr.}
\thanks{Davide Dardari is with DEI, CNIT at University of Bologna, Italy. Email: davide.dardari@unibo.it.}
\thanks{This work has been supported in part by the Belgian network IAP Bestcom and the EU network of excellence NEWCOM\#.}
}

\maketitle



\begin{abstract}
Threshold and ambiguity phenomena are studied in Part 1 of this work \cite{part1} where approximations for the mean-squared-error (MSE) of the maximum likelihood estimator are proposed using the method of interval estimation (MIE), and where approximate upper and lower bounds are derived.
In this part we consider time-of-arrival estimation and we employ the MIE to derive closed-form expressions of the begin-ambiguity, end-ambiguity and asymptotic signal-to-noise ratio (SNR) thresholds with respect to some features of the transmitted signal. Both baseband and passband pulses are considered.
We prove that the begin-ambiguity threshold depends only on the shape of the envelope of the ACR, whereas the end-ambiguity and asymptotic thresholds only on the shape of the ACR.
We exploit the results on the begin-ambiguity and asymptotic thresholds to optimize, with respect to the available SNR, the pulse that achieves the minimum attainable MSE.
The results of this paper are valid for various estimation problems.
\end{abstract}


\begin{IEEEkeywords}
	Nonlinear estimation, threshold and ambiguity phenomena, maximum likelihood	estimator, mean-squared-error, signal-to-noise ratio, time-of-arrival, optimal signal design.
\end{IEEEkeywords}

\IEEEpeerreviewmaketitle


\section{Introduction}\label{intro_sec}

\begin{figure}
  \centering
  \includegraphics[scale = 0.49]{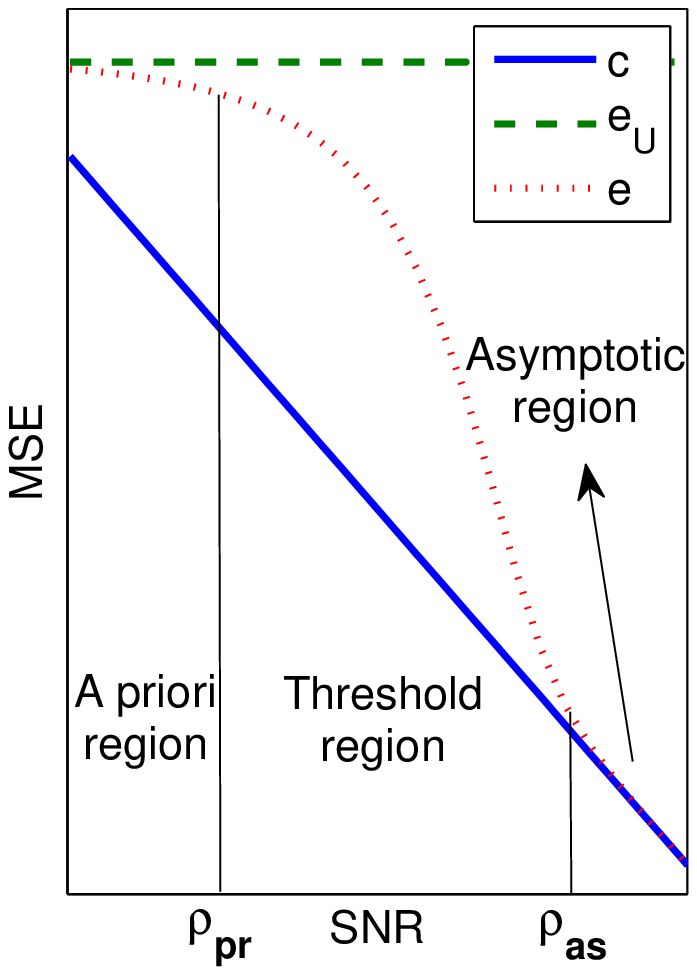}
  \includegraphics[scale = 0.49]{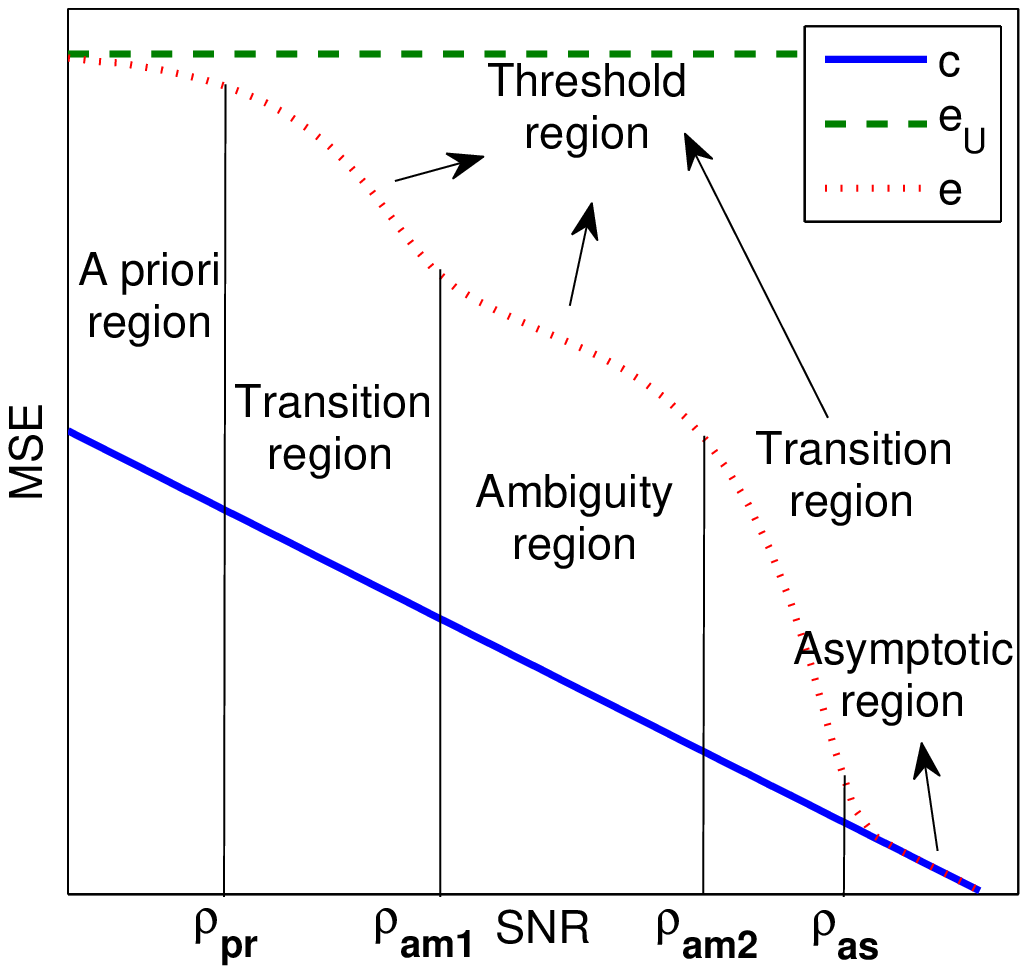}\\
  $\;\;\;\;\;$ (a) 
  $\;\;\;\;\;\;\;\;\;\;\;\;\;\;\;\;\;\;\;\;\;\;\;\;\;\;\;\;\;\;\;\;\;\;\;\;$ 
  (b) $\;\;\;\;\;\;\;\;$
  \caption{SNR regions (a) \textit{A priori}, threshold and asymptotic regions for non-oscillating ACR (b) \textit{A priori}, ambiguity and asymptotic regions for oscillating ACR ($c$: CRLB, $e_U$: MSE of uniform distribution in the \textit{a priori} domain, $e$: achievable MSE, $\rho_{pr}$, $\rho_{am1}$, $\rho_{am2}$, $\rho_{as}$: \textit{a priori}, begin-ambiguity, end-ambiguity and asymptotic thresholds).}
  \label{01_regions_pic}
\end{figure}

\IEEEPARstart{N}{onlinear} deterministic parameter estimation is subject to the threshold effect \cite{ziv,chow,weiss1,weiss2,zeira1,zeira2,sadler1,sadler2}. Due to this effect the signal-to-noise ratio (SNR) axis can be split into three regions as illustrated in Fig. \ref{01_regions_pic}(a):
\begin{enumerate}
	\item \textit{A priori} region: Region in which the estimator becomes uniformly distributed in the \textit{a priori} domain.
	\item Threshold region: Region of transition between the \textit{a priori} and asymptotic regions.
	\item Asymptotic region: Region in which an asymptotically efficient estimator, such as the maximum likelihood estimator (MLE), achieves the Cramer-Rao lower bound (CRLB). Otherwise, the estimator achieves its own asymptotic mean-squared-error (MSE) (e.g, MLE with random signals and finite snapshots \cite{Renaux2006,Renaux2007}).
\end{enumerate}

When the autocorrelation (ACR) with respect to (w.r.t.) the unknown parameter is oscillating, five regions can be identified as shown in Fig. \ref{01_regions_pic}(b):
1) the \textit{a priori} region, 2) the \textit{a priori}-ambiguity transition region, 3) the ambiguity region, 4) the ambiguity-asymptotic transition region, and 5) the asymptotic region. The MSE achieved in the ambiguity region is approximately equal to the envelope CRLB (ECRLB).
%
%
In Figs. \ref{01_regions_pic}(a) and \ref{01_regions_pic}(b), $\rho_{pr}$, $\rho_{am1}$, $\rho_{am2}$ and $\rho_{as}$, respectively, denote the \textit{a priori}, begin-ambiguity, end-ambiguity and asymptotic thresholds determining the limits of the defined regions.

\smallskip

As the evaluation of the statistics of most estimators such as the MLE is often unattainable in the threshold region, many lower bounds have been proposed \cite{VanBell2007,renaux} for both deterministic (the unknown parameter has only one possible value) and Bayesian (the unknown parameter follows a given \textit{a priori} distribution) estimation in order to be used as benchmarks and to describe the behavior of an estimator in that region. 

\smallskip

Threshold computation is considered in \cite{weiss1,weiss2} where the \textit{a priori}, begin-ambiguity, end-ambiguity and asymptotic thresholds are computed based on the Ziv-Zakai lower bound (ZZLB); the ZZLB evaluates accurately the asymptotic threshold and detects roughly the ambiguity region.
Thresholds are also computed in \cite{zeira1,zeira2} using the Barankin lower bound (BLB); the obtained thresholds are much smaller than the true ones.
Closed-form expressions of the asymptotic threshold are derived in \cite{Steinhardt} for frequency estimation and in \cite{Richmond2005} for angle estimation by employing the method of interval estimation (MIE). 
The method in \cite{Steinhardt} is based on the MSE approximation (MSEA) in \cite{rife} and is valid for cardinal sine ACRs only, whereas that in \cite{Richmond2005} is based on the probability of non-ambiguity and can be used with any ACR shape. The approaches in \cite{Steinhardt,Richmond2005} are discussed in details and compared to our approach in Sec. \ref{threshold_sec}.

\smallskip

Optimal power allocation for multicarrier systems with interference is considered in \cite{signaldesign}; the approach followed therein minimizes the CRLB for TOA estimation without taking into account the threshold and ambiguity effects.
Optimal pulse design for TOA estimation is studied in \cite{mcaulay3} based on the BLB; the authors study the reduction of the asymptotic threshold by considering different ACR shapes.
The optimization of the time-bandwidth product for frequency estimation is investigated in \cite{Van1968} based on the MIE. The approach in \cite{Van1968} is discussed and compared to ours in Sec. \ref{signaldesign_sec}.

\smallskip

In Part 1 of this work \cite{part1}, an approximate upper bound and various MSEAs for the MLE are proposed by using the MIE \cite{Wood1955,Kote1959,wozencraft,Van1968,mcaulay3,rife,Boyer2004,athley,Najjar2005,Richmond2005,Richmond2006,VanBell2007}. Some approximate lower bounds (ALB) are proposed as well by employing the binary detection principle first used by Ziv and Zakai \cite{ziv}.

\smallskip

In Part 2 (current paper), we utilize an MIE-MSEA (proposed later in Sec. \ref{interval_sec}) to derive analytic expressions of the begin-ambiguity, end-ambiguity and asymptotic thresholds.
The obtained thresholds are very accurate (in particular the end-ambiguity and asymptotic thresholds of oscillating ACRs).
To the best of our knowledge, our approach is the first utilizing an MIE-based MSEA (very accurate approximation) and that can be used with any ACR shape.
%
%
The equations established in this paper are obtained by considering TOA estimation. However, our method can be applied on any estimation problem satisfying the system model of Part 1.

\smallskip

We prove that the begin-ambiguity threshold only depends on the shape of the ACR envelope (e.g, cardinal sine, Gaussian, raised cosine with fixed roll-off) regardless of other parameters (e.g, \textit{a priori} domain, bandwidth, mean frequency), and the end-ambiguity and asymptotic thresholds only depend on the ACR shape (which can be described by the envelope shape and the mean frequency to bandwidth ratio, together) regardless of other parameters (e.g, the bandwidth and the mean frequency if their ratio is constant).
%
%
The thresholds of the different SNR regions are also evaluated numerically using an MSEA and two ALBs (derived in Part 1).
We show that the \textit{a priori} threshold depends on both the \textit{a priori} domain and the shape of the ACR envelope.

\smallskip

By making use of the obtained results about thresholds, we propose a method to optimize, w.r.t. the available SNR, the spectrum of the transmitted pulse in order to achieve the minimum attainable MSE. The proposed method is very simple and very accurate.
To the best of our knowledge, this is the first optimization problem addressing the minimization of the MSE subject to the threshold and ambiguity phenomena. 

\smallskip

The rest of the paper is organized as follows.
In Sec. \ref{model_sec} we describe the system model. 
In Sec. \ref{mseapprox_sec} we introduce some MIE-based MSEAs and ALBs.
In Sec. \ref{threshold_sec} we consider the numerical and analytical computation of the thresholds and analyze their properties. 
In Sec. \ref{thr_num_sec} we present and discuss some numerical results about the thresholds when baseband and passband pulses are employed.
In Sec. \ref{signaldesign_sec} we propose a method to optimize the spectrum of the transmitted pulse w.r.t. the available SNR.


\section{System model}\label{model_sec}

In this section we describe our system model.
Let $s(t)$ be the transmitted signal, $\alpha$ and $\Theta$ the positive gain and the time delay introduced by an additive white Gaussian noise (AWGN) channel, and $\tilde{w}(t)$ the noise with two-sided power spectral density (PSD) of $\frac{N_0}{2}$. We can write the received signal as:
\begin{equation}\nonumber
	r(t) = \alpha s(t-\Theta) + \tilde{w}(t).
\end{equation}
We assume that $\Theta$ is deterministic with $D_{\Theta}=[\Theta_1,\Theta_2]$ representing its \textit{a priori} domain.

\smallskip

From Part 1, the MLE of $\Theta$ is given by 
\begin{eqnarray}
	\hat{\Theta} = \argmax{\theta}{\left\{X_{r,s}(\theta)\right\}} \nonumber
\end{eqnarray}
where $X_{r,s}(\theta)=\alpha R_s(\theta-\Theta)+w(\theta)$ is the CCR of $r(t)$ and $s(t)$ with $R_s(\theta)=\int_{-\infty}^{+\infty}s(t)s(t-\theta)dt$ being the ACR of $s(t)$ and $w(\theta)=\int_{-\infty}^{+\infty}r(t)\tilde{w}(t-\theta)dt$ a zero-mean colored Gaussian noise of covariance $C_{w}(\theta)=\frac{N_0}{2}R_s(\theta).$

\smallskip

From Part 1, we can express the CRLB, the ECRLB and the maximum MSE of $\Theta$ as:
\begin{eqnarray}
	c &=& \frac{1}{\rho\beta_s^2} \label{crlb_toa_eq}\\
	c_e &=& \frac{1}{\rho\beta_e^2} \label{envcrlb_toa_eq}\\
	e_U &=& \frac{(\Theta_2-\Theta_1)^2}{12}+\Big[\Theta-\frac{\Theta_1+\Theta_2}{2}\Big]^2 \label{mseU_eq}
\end{eqnarray}
where $\rho = \frac{\alpha^2 E_s}{N_0/2}$ denotes the SNR, and $\beta_s^2$ and $\beta_e^2$ stand for the mean quadratic bandwidth (MQBW) and the envelope MQBW (EMQBW) of $s(t)$, respectively. We have:
\begin{align}
	\beta_s^2 &= -\frac{\ddot{R}_s(0)}{E_s} = \beta_e^2 + 4\pi^2f_c^2 \approx 4\pi^2f_c^2 \label{mqbw_toa_eq}
\end{align}
where $\ddot{R}_s(\theta)$ denotes the second derivative of $R_s(\theta)$, $E_s=\int_{-\infty}^{+\infty}s^2(t)dt$ and $f_c=\frac{\int_0^{+\infty}f|\mathcal{F}_s(f)|^2df}{\int_0^{+\infty}|\mathcal{F}_s(f)|^2df}$ represent the energy and the mean frequency of $s(t)$, with $\mathcal{F}_s(f)$ being the Fourier transform of $s(t)$.

\smallskip

We have seen in Part 1, that for a signal occupying the whole band from 3.1 to 10.6 GHz\footnote{The ultra wideband (UWB) spectrum authorized for unlicensed use by the US federal commission of communications in May 2002 \cite{fcc}.} ($f_c=6.85$ GHz, bandwidth $B=7.5$ GHz), we have $\beta_e^2=\frac{\pi^2B^2}{3}\approx\frac{4\pi^2f_c^2}{10}$, so $c\approx\frac{c_e}{11}$. Therefore, the estimation performance seriously deteriorates if the ECRLB is achieved instead of the CRLB due to ambiguity.

\smallskip

As $\beta_e^2<<4\pi^2f_c^2$, the super accuracy associated with $c$ is mainly due to the mean frequency $f_c$.
To benefit from this super accuracy at sufficiently high SNRs, the sufficient condition to satisfy is that the phase of the transmitted signal should not be modified across the channel (e.g, due to fading), regardless whether the signal is pure impulse-radio UWB (carrier-less), carrier-modulated with known phase (e.g, in monostatic radar), or carrier-modulated with unknown phase (e.g, in most communication systems). With the latter, we have to use the time difference of arrival (TDOA) technique.


\section{MSEAs and ALBs}\label{mseapprox_sec}

In this section we introduce some MSEAs and ALBs that will be used later in Secs. \ref{threshold_sec} and \ref{thr_num_sec} to compute the thresholds.


\subsection{MIE-based MSEAs}\label{interval_sec}

We have seen in Part 1 that by splitting the \textit{a priori} domain of $\Theta$ into $N$ intervals $D_n=[d_n,d_{n+1})$, $(n=n_1,\cdots,n_N)$, ($n_1\leq0$, $n_N\geq0$), we can write the MSE of $\hat{\Theta}$ as:
\begin{equation} \label{mse_eq}
	e(\rho) = \sum_{n=n_1}^{n_N}P_n\left[\left(\Theta-\mu_n\right)^2+\sigma^2_n\right]
\end{equation}
where $P_n = \mathbb{P}\{\hat{\Theta}\in D_n\}$ denotes the interval probability, and $\mu_n=\mathbb{E}\{\hat{\Theta}_n\}$ and $\sigma_n^2=\mathbb{E}\{(\hat{\Theta}_n-\mu_n)^2\}$ represent, respectively, the mean and the variance of the interval MLE  $\hat{\Theta}_n=\hat{\Theta}|\hat{\Theta}\in D_n$ ($\mathbb{P}$ and $\mathbb{E}$ stand for the probability and expectation operators).
For oscillating (resp. non-oscillating) ACRs, we consider an interval around each local maximum (resp. split $D_{\Theta}$ into $N$ equal duration intervals); $D_0$ always contains the maximum of the ACR.

\smallskip

Different approximations of $P_n$, $\mu_n$ and $\sigma^2_n$ were proposed in Part 1. Below, we only present the approximations that will be used later in this paper for the numerical and the analytic evaluation of the thresholds.

\smallskip


\subsubsection{An MSEA for numerical threshold computation}


We present in this paragraph the MSEA 
\begin{equation} \label{enum_eq}
	e_{\text{num}}(\rho)
\end{equation}
based on \eqref{mse_eq} and that we will use later in Sec. \ref{thr_num_sec} for the numerical evaluation of the different thresholds; $e_{\text{num}}(\rho)$ is the most accurate approximation proposed in Part 1.

\smallskip

For both oscillating and non-oscillating ACRs, $P_n$ in \eqref{mse_eq} is approximated by $P^{(1)}_n = \text{GenzAlgo}(\theta_{n_1},\cdots,\theta_{n_N})$
where $\text{GenzAlgo}$ denotes one of Genz's algorithms written based on \cite{genz1,genz2,genz3,genz4} to compute the multivariate normal probability with integration region specified by a set of linear inequalities (see Part 1 for more details), and $\theta_n$ represents a testpoint in $D_n$; 
$\theta_n$ is selected as the abscissa of the $n$th local maximum (resp. the center of $D_n$) for oscillating (resp. non-oscillating) ACRs; 
$\theta_0=\Theta$ (abscissa of the maximum) for both ACR types.

\smallskip

For oscillating (resp. non-oscillating) ACRs, $\mu_n$ and $\sigma^2_n$ are approximated by 
$\mu_{n,1,o}=\theta_n$ and $\sigma^2_{n,1,o}=\min\left\{c\frac{\ddot{R}_0^2}{\ddot{R}_n^2},\sigma^2_{n,U}\right\}$ 
(resp. $\mu_{n,1,c}=d_nP\{d_n\}+d_{n+1}P\{d_{n+1}\}$ and $\sigma^2_{n,1,c}=\min\left\{\sigma^2_{n,B},\sigma^2_{n,U}\right\}$) 
where $\ddot{R}_n=\left.\frac{d^2R_s(\theta)}{d\theta^2}\right|_{\theta=\theta_n}$, 
$\sigma^2_{n,U} = \frac{(d_{n+1}-d_n)^2}{12}$, $P\{d_n\}=Q\left(\sqrt{\rho}\frac{\dot{R}_n}{E_s\beta_s}\right)$ 
and $\sigma^2_{n,B}=\mathbb{P}\{d_n\}(1-\mathbb{P}\{d_n\})(d_{n+1}-d_n)^2$, 
with $Q(y)=\frac{1}{\sqrt{2\pi}}\int_{y}^\infty e^{-\frac{\xi^2}{2}}d\xi$ being the Q function
and $\dot{R}_n=\left.\frac{dR_s(\theta)}{d\theta}\right|_{\theta=\theta_n}$.

\smallskip


\subsubsection{An MSEA for analytic threshold computation}

The MSEA $e_{\text{ana}}(\rho)$ proposed in this paragraph will be used later in Sec. \ref{thr_anal_sec} to express analytically the end-ambiguity and asymptotic thresholds; $e_{\text{ana}}(\rho)$ employs the probability upper bound proposed by McAulay in \cite{mcaulay3}. It evaluates the achieved MSE in the intervals $D_{-1}$, $D_0$ and $D_1$, which means that the SNR is supposed to be relatively high.

\smallskip

By approximating $\mu_n$ in \eqref{mse_eq} by $\theta_n$, 
approximating $\sigma^2_0$ by $c$, 
neglecting $\sigma^2_{\pm1}$ ($\sigma^2_{\pm1}<<(\Theta-\mu_{\pm1})^2$), 
taking $\theta_0=\Theta$ and $\theta_{\pm1}=\Theta\pm\Delta$ 
with $\Delta=\frac{1}{f_c}\approx\frac{2\pi}{\beta_s}$ for oscillating ACRs ($\theta_{\pm1}$ are the approximate abscissa of the two local maxima around the global one) 
and $\Delta=\frac{\pi}{4\beta_s}$ for non-oscillating ACRs ($\theta_{\pm1}$ are empirically chosen, see Sec. V-B in Part 1 for more details), 
and adopting the McAulay probability upper bounds $P^{(2)}_0=1$ and $P^{(2)}_{\pm1}=Q\left(\sqrt{\frac{\rho}{2}\left[1-R(\Delta)\right]}\right)$ 
with $R(\theta)=\frac{R_s(\theta)}{E_s}$ denoting the normalized ACR,
$e_{\text{ana}}(\rho)$ becomes
%
%
\begin{equation} \label{eana_eq}
	e_{\text{ana}}(\rho) = c + 2\Delta^2Q\left(\sqrt{\frac{\rho}{2}[1-R(\Delta)]}\right).
\end{equation}

Let us now explain why $e_{\text{ana}}(\rho)$ is appropriate for the evaluation of the end-ambiguity and asymptotic thresholds.
Assume for the moment that the CRLB is achieved (i.e. the SNR is sufficiently high). In the course of decreasing the SNR, the threshold (resp. ambiguity) region begins for non-oscillating (resp. oscillating) ACRs when the estimates of the unknown parameter start to spread along the ACR (resp. the local maxima of the ACR) instead of falling in the vicinity of the maximum (resp. global maximum). Therefore, the estimates only fall at the end of the threshold and ambiguity regions (if we start from low SNRs) in the interval $D_0$ and the intervals $D_{-1}$ and $D_1$ (at the left and the right of $D_0$) so the achieved MSE can be approximated using $e_{\text{ana}}(\rho)$.


\subsection{Binary detection based ALBs}\label{zivzakai_sec}

By using the principle of binary detection, we have derived in Part 1 the following ALBs ($i=1,2$):
\begin{eqnarray}
	z_i &=& \int_{0}^{\epsilon_i} \xi Q\left(\sqrt{\frac{\rho}{2}[1-R(\xi)]}\right)d\xi \label{z0_eq}\\
	b_i &=& \int_{0}^{\epsilon_i} \xi V\left\{Q\left(\sqrt{\frac{\rho}{2}[1-R(\xi)]}\right)\right\}d\xi \label{b0_eq}
\end{eqnarray}
where $\epsilon_1=\min\{\Theta-\Theta_1,2(\Theta_2-\Theta)\}$ and $\epsilon_2=\min\{\Theta_2-\Theta,2(\Theta-\Theta_1)\}$;
$V\{f(\xi)\} = \max\{f(\zeta\geq\xi)\}$ denotes the valley-filling function.
%
%
We have seen in Part 1 that $z_i$ and $b_i$ are very tight and that $b_i$ is tighter than $z_i$; $z_1$ and $b_1$ are, respectively, tighter than $z_2$ and $b_2$ when $\theta_0-\Theta_1>\Theta_2-\theta_0$. 


\section{Threshold computation}\label{threshold_sec}

We consider in this section the computation of the thresholds of the different SNR regions w.r.t. some features of the transmitted signal.

\smallskip

Similarly to Part 1, we define the \textit{a priori} $\rho_{pr}$, begin-ambiguity $\rho_{am1}$, end-ambiguity $\rho_{am2}$ and asymptotic $\rho_{as}$ thresholds as \cite{weiss2}:
\begin{eqnarray}
	\rho_{pr} &=& \rho \; : \; e(\rho)=\alpha_{pr}e_U \label{thpr_eq}\\
	\rho_{am1} &=& \rho \; : \; e(\rho)=\alpha_{am1}c_e \label{tham1_eq}\\
	\rho_{am2} &=& \rho \; : \; e(\rho)=\alpha_{am2}c_e \label{tham2_eq}\\
	\rho_{as} &=& \rho \; : \; e(\rho)=\alpha_{as}c. \label{thas_eq}
\end{eqnarray}
%
%
We take $\alpha_{pr}=0.5$, $\alpha_{am1}=2$, $\alpha_{am2}=0.5$ and $\alpha_{as}=1.1$.

\smallskip

The considered features of the transmitted signal are the \textit{a priori} time bandwidth product (ATBW) and the inverse fractional bandwidth (IFBW) defined as:
\begin{eqnarray}
	\gamma &=& TB \label{atbw_eq}\\
	\lambda &=& \frac{f_c}{B} \label{ifbw_eq}
\end{eqnarray}
where $T=\Theta_2-\Theta_1$ (\textit{a priori} time) is the width of the \textit{a priori} domain of $\Theta$ and $B$ the bandwidth of the transmitted signal.

\smallskip

In Sec. \ref{threshold_numcomput_sec}, we consider the numerical calculation of the thresholds.
We derive in Sec. \ref{thr_anal_sec} analytic expressions of the begin-ambiguity, end-ambiguity and asymptotic thresholds,
and discuss in Sec. \ref{thr_prop_sec} the properties of the thresholds obtained in Sec. \ref{thr_anal_sec}.


\subsection{Numerical computation}\label{threshold_numcomput_sec}

As mentioned above we consider here the numerical computation of the thresholds.
To find $\rho_{pr}$, $\rho_{am1}$, $\rho_{am2}$ and $\rho_{as}$ w.r.t. $\gamma$ (resp. $\lambda$) numerically, we vary $\gamma$ (resp. $\lambda$) by fixing $T$ (resp. $f_c$) and varying $B$ (or vice versa) and compute for each value of $\gamma$ (resp. $\lambda$) the achieved MSE along the SNR axis.
Then, the thresholds are then obtained by making use of \eqref{thpr_eq}, \eqref{tham1_eq}, \eqref{tham2_eq} and \eqref{thas_eq}.

\smallskip

Theoretically, the thresholds should be computed from the MSE achieved in practice. As the exact expression of the MSE is not obtainable in most estimation problems, the thresholds can be calculated using a MSEA, an upper bound or a lower bound.
In Sec. \ref{thr_num_sec}, the \textit{a priori}, begin-ambiguity and end-ambiguity thresholds are computed numerically using the MSEA $e_{\text{num}}(\rho)$ in \eqref{enum_eq}. The asymptotic threshold is computed using $e_{\text{num}}(\rho)$ and the ALBs $z_i$ in \eqref{z0_eq} and $b_i$ in \eqref{b0_eq}.


\subsection{Analytic expressions of the begin-ambiguity, end-ambiguity and asymptotic thresholds}\label{thr_anal_sec}

In this subsection, we derive analytic expressions of the begin-ambiguity, end-ambiguity and asymptotic thresholds by making use of the MSEA $e_{\text{ana}}(\rho)$ in \eqref{eana_eq}.

\smallskip


\subsubsection{Asymptotic threshold for oscillating and non-oscillating ACRs}\label{asthr_sec}

Let:
\begin{eqnarray}
	G(\rho) = \rho Q\left(\sqrt{\frac{\rho}{2}[1-R(\Delta)]}\right) \label{G_eq}
\end{eqnarray}
Using \eqref{crlb_toa_eq}, \eqref{eana_eq} and \eqref{G_eq} we can write from the asymptotic threshold definition in \eqref{thas_eq}:
\begin{equation}
	G(\rho_{as}) = G_{as} \label{Gas_eq}
\end{equation}
where
\begin{eqnarray}
	G_{as} = \frac{\alpha_{as}-1}{2\Delta^2\beta_s^2} \label{GasCte_eq}
\end{eqnarray}
denotes a constant; $\rho_{as}$ is the solution of \eqref{Gas_eq}.

\smallskip

To find an analytic expression of $\rho_{as}$ we consider the following approximation of the Q function
\begin{equation} \label{Qapprox_eq}
				Q(\xi) \approx \frac{1}{\xi}\frac{1}{\sqrt{2\pi}}e^{-\frac{\xi^2}{2}} , \xi>>1
\end{equation}
obtained from the inequality $\left(\frac{1}{\xi}-\frac{1}{\xi^3}\right)\frac{1}{\sqrt{2\pi}}e^{-\frac{\xi^2}{2}}<Q(\xi)<\frac{1}{\xi}\frac{1}{\sqrt{2\pi}}e^{-\frac{\xi^2}{2}}$, $\xi>0$ in \cite[pp. 83]{wozencraft}.
Let:
\begin{eqnarray}
	H(\rho) = -\frac{\rho[1-R(\Delta)]}{2} \label{H_eq}
\end{eqnarray}
From \eqref{GasCte_eq}, \eqref{Qapprox_eq} and \eqref{H_eq}, we can write \eqref{Gas_eq} as:
\begin{eqnarray}
	H(\rho_{as})e^{H(\rho_{as})} = H_{as} \label{Has_eq}
\end{eqnarray}
with
\begin{eqnarray}
	H_{as} = -\frac{\pi G_{as}^2[1-R(\Delta)]}{2} = -\frac{\pi(\alpha_{as}-1)^2[1-R(\Delta)]}{8\Delta^4\beta_s^4} \label{HasCte_eq}
\end{eqnarray}
so the asymptotic threshold in \eqref{Has_eq} can be expressed as:
\begin{eqnarray}
	\rho_{as} = \frac{-2W_{-1}(H_{as})}{1-R(\Delta)} \label{thasAnal_eq}
\end{eqnarray}
where $W_{-1}(\xi)$ denotes the branch ``$-1$" (because $H_{as}$ is negative) of the Lambert W function defined as a solution (more than one solution may exist) of the equation $We^W=\xi$. 
Like the other non-elementary functions (e.g, Q function, error function), the Lambert W function has Taylor series expansion and can be computed recursively; it is also implemented in MATLAB; hence, the solution in \eqref{H_eq} can be considered as an analytic solution since it can directly be obtained.

\smallskip

We recall that in the evaluation of $G_{as}$ in \eqref{GasCte_eq}, $H_{as}$ in \eqref{HasCte_eq} and $\rho_{as}$ in \eqref{thasAnal_eq}, we take $\Delta=\frac{\pi}{4\beta_s}$ for non-oscillating ACRs and $\Delta\approx\frac{1}{f_c}\approx\frac{2\pi}{\beta_s}$ for oscillating ACRs.

\smallskip


\subsubsection{End-ambiguity threshold for oscillating ACRs}\label{am2thr_sec}

From the end-ambiguity threshold definition in \eqref{tham2_eq} we can write using \eqref{crlb_toa_eq}, \eqref{envcrlb_toa_eq}, \eqref{mqbw_toa_eq}, \eqref{eana_eq} and \eqref{G_eq}:
\begin{eqnarray}
	G(\rho_{am2}) = G_{am2} \label{Gam2_eq}
\end{eqnarray}
where 
\begin{eqnarray}
	G_{am2} = \frac{1}{2\Delta^2}\left(\frac{\alpha_{am2}}{\beta_e^2}-\frac{1}{\beta_s^2}\right) 
				\approx \frac{\alpha_{am2}}{2\Delta^2\beta_e^2}. \label{Gam2Cte_eq}
\end{eqnarray}
Using \eqref{Qapprox_eq}, \eqref{H_eq} and \eqref{Gam2Cte_eq}, we can write \eqref{Gam2_eq} as:
\begin{eqnarray}
	H(\rho_{am2})e^{H(\rho_{am2})} = H_{am2} \label{Ham2_eq}
\end{eqnarray}
where
\begin{eqnarray}
	H_{am2} = -\frac{\pi G_{am2}^2[1-R(\Delta)]}{2} 
				\approx -\frac{\pi\alpha_{am2}^2[1-R(\Delta)]}{8\Delta^4\beta_e^4} \label{Ham2Cte_eq}
\end{eqnarray}
so the end-ambiguity threshold in \eqref{Ham2_eq} can be expressed as:
\begin{eqnarray}
	\rho_{am2} = \frac{-2W_{-1}(H_{am2})}{1-R(\Delta)}. \label{tham2Anal_eq}
\end{eqnarray}
We recall that in the evaluation of $G_{am2}$ in \eqref{Gam2Cte_eq}, $H_{am2}$ in \eqref{Ham2Cte_eq} and $\rho_{am2}$ in \eqref{tham2Anal_eq}, we take $\Delta\approx\frac{1}{f_c}\approx\frac{2\pi}{\beta_s}$.

\smallskip


\subsubsection{Begin-ambiguity threshold for oscillating ACRs}\label{am1thr_sec}

To compute the begin-ambiguity threshold, we cannot employ the MSEA in \eqref{eana_eq} because the estimates fall now, not only in $D_{-1}$, $D_0$ and $D_1$, but around all the local maxima in the vicinity of the maximum of the envelope of the ACR.
Therefore, by considering the envelope $e_R(\theta)$ of the normalized ACR $R(\theta)$ instead of $R(\theta)$ itself, and the ECRLB $c_e$ in \eqref{envcrlb_toa_eq} instead of the CRLB $c$ in \eqref{crlb_toa_eq}, we can approximate the MSE in the vicinity of the maximum of $e_R(\theta)$ by:
\begin{align}
	e_{\text{ana},e}(\rho) \approx c_e + 2\Delta^2Q\left(\sqrt{\frac{\rho}{2}[1-e_R(\Delta)]}\right) \label{eana_env_eq} 
\end{align}
where, similarly to the case of non-oscillating ACRs, we take $\Delta=\frac{\pi}{4\beta_e}$ ($\beta_s$ is replaced by $\beta_e$ because the EMQBW is equal to the MQBW of the envelope).
Let:
\begin{eqnarray}
	G_e(\rho) &=& \rho Q\left(\sqrt{\frac{\rho}{2}[1-e_R(\Delta)]}\right) \label{Ge_eq}\\
	H_e(\rho) &=& -\frac{\rho[1-e_R(\Delta)]}{2} \label{He_eq}
\end{eqnarray}
From \eqref{envcrlb_toa_eq}, \eqref{eana_env_eq}, \eqref{Ge_eq} and \eqref{He_eq} we can write the definition of the begin-ambiguity threshold in \eqref{tham1_eq} as:
\begin{equation} \label{Gam1_eq}
	G_e(\rho_{am1}) = G_{am1}
\end{equation}
where
\begin{equation} \label{Gam1Cte_eq}
	G_{am1} = \frac{\alpha_{am1}-1}{2\Delta^2\beta_e^2}.
\end{equation}
%
Using \eqref{Qapprox_eq}, \eqref{Gam1_eq} becomes:
\begin{equation} \label{Ham1_eq}
	H_e(\rho_{am1})e^{H_e(\rho_{am1})} = H_{am1}
\end{equation}
where
\begin{equation} \label{Ham1Cte_eq}
	\textstyle H_{am1} = -\frac{\pi G_{am1}^2[1-e_R(\Delta)]}{2} 
				= -\frac{\pi(\alpha_{am1}-1)^2[1-e_R(\Delta)]}{8\Delta^4\beta_e^4}
\end{equation}
so we can express the begin-ambiguity threshold from \eqref{Ham1_eq} as:
\begin{eqnarray}
	\rho_{am1} = \frac{-2W_{-1}(H_{am1})}{1-e_R(\Delta)}. \label{tham1Anal_eq}
\end{eqnarray}
We recall that in the evaluation of $G_{am1}$ in \eqref{Gam1Cte_eq}, $H_{am1}$ in \eqref{Ham1Cte_eq} and $\rho_{am1}$ in \eqref{tham1Anal_eq}, we take $\Delta=\frac{\pi}{4\beta_e}$.

\smallskip


\subsubsection{About the end-ambiguity and asymptotic thresholds for oscillating ACRs}\label{asam2thr_sec}

Note that in the computation of the end-ambiguity and asymptotic thresholds for oscillating ACRs, $R(\Delta)$ can be replaced by $e_R(\Delta)$ because $\theta_{\pm1}$ in \eqref{eana_eq} are the abscissa of two local maxima of $R(\theta-\Theta)$ (the local maxima are located on the envelope).
Therefore, $\rho_{as}$ in \eqref{thasAnal_eq} and $\rho_{am2}$ in \eqref{tham2Anal_eq} can be expressed as:
\begin{eqnarray}
	\rho_{as} &=& \frac{-2W_{-1}(H_{as})}{1-e_R(\Delta)} \label{thasAnal1_eq}\\
	\rho_{am2} &=& \frac{-2W_{-1}(H_{am2})}{1-e_R(\Delta)} \label{tham2Anal1_eq}
\end{eqnarray}
where
\begin{eqnarray}
	H_{as} &=& -\frac{\pi(\alpha_{as}-1)^2[1-e_R(\Delta)]}{8\Delta^4\beta_s^4} \label{HasCte1_eq}\\
	H_{am2} &=& -\frac{\pi\alpha_{am2}^2[1-e_R(\Delta)]}{8\Delta^4\beta_e^4}. \label{Ham2Cte1_eq}
\end{eqnarray}
By using \eqref{thasAnal1_eq} and \eqref{tham2Anal1_eq} instead of \eqref{thasAnal_eq} and \eqref{tham2Anal_eq}, we highly simplify the calculation of the thresholds. 
In fact, if we want to compute the thresholds of a passband pulse (i.e. pulse modulated by carrier) w.r.t. the IFBW $\lambda$ in \eqref{ifbw_eq}, then instead of generating the normalized ACR $R(\theta)$ for each value of $\lambda$, we just compute the normalized ACR envelope $e_R(\theta)$ once and evaluate $R(\Delta)=e_R(\Delta)$ by varying $\Delta$ w.r.t. $\lambda$.


\subsection{Threshold properties}\label{thr_prop_sec}

In this subsection we prove that for a baseband (i.e. unmodulated) pulse that can be written as (e.g, Gaussian, cardinal sine and raised cosine pulses):
\begin{equation} \label{wB_eq}
	w_B(t) = w_1(t') ,\; t'=Bt
\end{equation}
with $B$ denoting the bandwidth, the asymptotic threshold only depends on the shape $w_1(t)$ (i.e. independent of $B$) (e.g, constant for Gaussian and cardinal sine pulses, and function of the roll-off factor for raised cosine pulses), and that for the passband pulse
\begin{align} \label{wBfc_eq}
	w_{B,f_c}(t) &= w_B(t)\cos(2\pi f_ct) \nonumber\\
				&= w_1(t')\cos(2\pi\lambda t') ,\; t'=Bt
\end{align}
with $f_c$ denoting the carrier frequency, the begin-ambiguity threshold only depends on the shape $w_1(t)$ of the envelope $w_B(t)$ of $w_{B,f_c}(t)$ (i.e. independent of $B$, $f_c$ and the IFBW $\lambda$), whereas the end-ambiguity and asymptotic thresholds are functions of the shape $w_1(t)$ and the IFBW $\lambda$ in \eqref{ifbw_eq} (i.e. independent of the values taken by $B$ and $f_c$ separately). 
This is equivalent to saying that the begin-ambiguity threshold is only function of the shape of the envelope of the signal, whereas the end-ambiguity and asymptotic thresholds are only functions of the shape of the signal itself, regardless of any other parameters like the bandwidth and the carrier.

\smallskip


\subsubsection{Asymptotic threshold for baseband pulses}\label{PropAsBB_sec}

Let us prove that the asymptotic threshold in \eqref{thasAnal_eq} of the pulse $w_B(t)$ in \eqref{wB_eq} is independent of $B$. From \eqref{wB_eq} we can write the normalized ACR $R_B(\theta)$ of $w_B(t)$ as:
\begin{align}
	R_B(\theta) &= \textstyle \frac{\int_{-\infty}^{+\infty}w_B(t)w_B(t-\theta)dt}{\int_{-\infty}^{+\infty}w_B^2(t)dt}
				= \frac{\int_{-\infty}^{+\infty}w_1(t')w_1(t'-\theta')dt'}{\int_{-\infty}^{+\infty}w_1^2(t')dt'} \nonumber\\
				&= R_1(\theta') ,\; \theta'=B\theta \label{RB_eq}
\end{align}
where $R_1(\theta)$ denotes the normalized ACR of $w_1(t)$,
and the MQBW $\beta_B^2$ of $w_B(t)$ using \eqref{mqbw_toa_eq} and \eqref{RB_eq} as:
\begin{align}
	\beta_B^2 &= \left.-\frac{d^2R_B(\theta)}{d\theta^2}\right|_{\theta=0} 
				= \left.-B^2\frac{d^2R_1(\theta')}{d\theta'^2}\right|_{\theta'=0}
				= B^2\beta_1^2 \label{beta2B_eq}
\end{align}
where $\beta_1^2=-\ddot{R}_1(0)$ denotes the MQBW of $w_1(t)$ (unitary MQBW, i.e. MQBW per a bandwidth of $B=1$ Hz).
Note that $R_B(\theta)$ and $\beta_B$ used here are, respectively, equivalent to $R(\theta)$ and $\beta_s$ used in Sec. \ref{thr_anal_sec}.
As $\Delta=\frac{\pi}{4\beta_s}=\frac{\pi}{4\beta_B}$ for non-oscillating ACRs, we can write $R(\Delta)$ and $H_{as}$ in \eqref{thasAnal_eq} from \eqref{RB_eq} and \eqref{beta2B_eq} as:
\begin{eqnarray}
	R(\Delta) &=& R_B\left(\frac{\pi}{4\beta_B}\right) = R_B\left(\frac{\pi}{4B\beta_1}\right) 
				= R_1\left(\frac{\pi}{4\beta_1}\right) \nonumber\\
	H_{as} &=& -\frac{32(\alpha_{as}-1)^2\left[1-R_1\left(\frac{\pi}{4\beta_1}\right)\right]}{\pi^3}. \nonumber
\end{eqnarray}
We can see that both $R(\Delta)$ and $H_{as}$ are independent of $B$.
Hence, for the pulse in \eqref{wB_eq} the asymptotic threshold is independent of $B$; it depends only on the shape of the normalized ACR $R_B(\theta)$ determined by $R_1(\theta)$.

\smallskip


\subsubsection{Begin-ambiguity threshold for passband pulses}\label{PropAm1PB_sec}

Let us prove that the begin-ambiguity threshold in \eqref{tham1Anal_eq} of the pulse $w_{B,f_c}(t)$ in \eqref{wBfc_eq} is independent of $B$ and $f_c$.
The envelope $e_{R_{B,f_c}}(\theta)$ of the normalized ACR $R_{B,f_c}(\theta)$ of $w_{B,f_c}(t)$ and the EMQBW $\beta_{e,B,f_c}^2$ of $w_{B,f_c}(t)$ can be written from \eqref{wBfc_eq}, \eqref{RB_eq} and \eqref{beta2B_eq} as:
\begin{eqnarray}
	e_{R_{B,f_c}}(\theta) &=& R_B(\theta) = R_1(\theta') ,\; \theta'=B\theta \label{RBfc_eq}\\
	\beta_{e,B,f_c}^2 &=& \beta_B^2 = B^2\beta_1^2. \label{beta2eBfc_eq}
\end{eqnarray}
Note that $e_{R_{B,f_c}}(\theta)$ and $\beta_{e,B,f_c}^2$ used here are, respectively, equivalent to $e_R(\theta)$ and $\beta_e$ used in Sec. \ref{thr_anal_sec}.
As $\Delta=\frac{\pi}{4\beta_e}=\frac{\pi}{4\beta_{e,B,f_c}}$ for the begin-ambiguity threshold, we can write $e_R(\Delta)$ and $H_{am1}$ in \eqref{tham1Anal_eq} using \eqref{RBfc_eq} and \eqref{beta2eBfc_eq} as:
\begin{eqnarray}
	e_R(\Delta) &=& R_B\left(\frac{\pi}{4\beta_B}\right) = R_1\left(\frac{\pi}{4\beta_1}\right) \nonumber\\
	H_{am1} &=& -\frac{32(\alpha_{am1}-1)^2\left[1-R_1\left(\frac{\pi}{4\beta_1}\right)\right]}{\pi^3}. \nonumber
\end{eqnarray}
Both $e_R(\Delta)$ and $H_{am1}$ are independent of $B$ and $f_c$.
Hence, for the pulse in \eqref{wBfc_eq} the begin-ambiguity threshold is independent of $B$ and $f_c$; it only depends on the shape $R_1(\theta)$ of the envelope $e_{R_{B,f_c}}(\theta)$ of the normalized ACR $R_{B,f_c}(\theta)$.

\smallskip


\subsubsection{End-ambiguity and asymptotic thresholds for passband pulses}\label{PropAm2AsPB_sec}

Let us prove that the asymptotic threshold in \eqref{thasAnal1_eq} and the end-ambiguity threshold in \eqref{tham2Anal1_eq} of the pulse $w_{B,f_c}(t)$ in \eqref{wBfc_eq} are function of the shape $w_1(t)$ of the envelope $w_B(t)$ in \eqref{wB_eq} and the IFBW $\lambda$ in \eqref{ifbw_eq} only.

\smallskip

As $\Delta\approx\frac{1}{f_c}\approx\frac{2\pi}{\beta_s}$ for oscillating ACRs, we can write $e_R(\Delta)$, $H_{as}$ and $H_{am2}$ in \eqref{thasAnal1_eq} and \eqref{tham2Anal1_eq} using \eqref{RBfc_eq} and \eqref{beta2eBfc_eq} as:
\begin{eqnarray}
	e_R(\Delta) &=& R_B\left(\frac{1}{f_c}\right) = R_1\left(\frac{1}{\lambda}\right) \nonumber\\
	H_{as} &=& -\frac{(\alpha_{as}-1)^2[1-R_1\left(\frac{1}{\lambda}\right)]}{128\pi^3} \nonumber\\
	H_{am2} &=& -\frac{\pi\alpha_{am2}^2\lambda^4[1-R_1\left(\frac{1}{\lambda}\right)]}{8\beta_1^4}. \nonumber
\end{eqnarray}
Hence, the end-ambiguity and asymptotic thresholds of $w_{B,f_c}(t)$ are independent of $B$ and $f_c$ separately; they depend on the shape $R_1(\theta)$ of the envelope of the ACR and on the IFBW $\lambda$. Note that $R_1(\theta)$ and $\lambda$ determine together the shape of the ACR of $w_{B,f_c}(t)$.

\smallskip

We have mentioned in Sec. \ref{intro_sec} that a closed-form expression of the asymptotic threshold is derived in \cite{Steinhardt} based on the MIE-based MSEA in \cite{rife}. The obtained result is very nice. However, it is only applicable on cardinal sine ACRs. Furthermore, the employed MSEA considers the unknown parameter and the zeros of the ACR as testpoints. This choice is not optimal for asymptotic threshold computation because the MSE starts to deviate from the asymptotic MSE (the CRLB for asymptotically estimators) when the estimate starts to fall around the strongest local maxima.

\smallskip

The latter problem is bypassed in \cite{Richmond2005} by only considering the unknown parameter and the two strongest local maxima (like in our approach). However, the threshold is not computed based on the achieved MSE w.r.t. the asymptotic one (like in the approach of \cite{Steinhardt} and ours) but based on the probability of non-ambiguity. Obviously, the MSE-based approach is more reliable because the main concern in estimation is to minimize the MSE (by making it attaining the asymptotic one).

\smallskip

In this section we have two main contributions. The first is that we derived closed-from expressions of the begin-ambiguity, end-ambiguity and asymptotic thresholds for oscillating and non-oscillating ACRs. The obtained thresholds are very accurate (especially for the end-ambiguity and asymptotic thresholds of oscillating ACRs, see Sec. \ref{thr_num_sec}). Our approach can be applied on any estimation problem satisfying the system model of Part 1.
To the best of our knowledge, our results are completely new.
Also, we have dealt with the case of non-oscillating ACRs. To the best of our knowledge, no one has investigated this case before.

\smallskip

The second contribution is that we proved some properties of the obtained thresholds. The proved properties are valid for any estimation problem whose ACR (rather than transmitted signal like in the TOA case) satisfies \eqref{wB_eq} and \eqref{wBfc_eq}.


\section{Numerical results about thresholds}\label{thr_num_sec}

In this section we discuss some numerical results about the thresholds obtained for the baseband and passband Gaussian pulses respectively given by
\begin{eqnarray}
	g_{T_w}(t) &=& e^{-2\pi\frac{t^2}{T_w^2}} \label{unmodpulse_eq}\\
	g_{T_w,f_c}(t) &=& e_s(t)\cos(2\pi f_ct). \label{modpulse_eq}
\end{eqnarray}
The bandwidth at -10 dB of both $g_{T_w}(t)$ and $g_{T_w,f_c}(t)$ and the MQBW of $g_{T_w}(t)$ (equal to the EMQBW of $g_{T_w,f_c}(t)$) can respectively be expressed as \cite{dardari2}:
\begin{eqnarray}
	B &=& 2\sqrt{\frac{\ln10}{\pi}}\frac{1}{T_w} \label{B_gauss_eq}\\
	\beta^2 &=& \frac{2\pi}{T_w^2}. \label{envmqbw_gauss_eq}
\end{eqnarray}


\begin{figure}
  \centering
  \includegraphics[width=8cm]{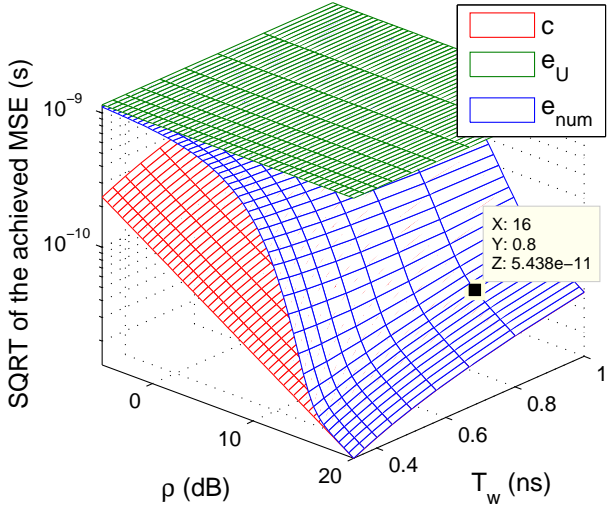}
  \caption{Baseband: SQRTs of the CRLB $c$, the maximum MSE $e_U$ and the MSEA $e_{\text{num}}$ w.r.t. the SNR $\rho$ and the pulse width $T_w$.}
  \label{07_varBB_snr_Tw_pic}
\end{figure}

\smallskip

In Sec. \ref{07_01_threshold_sec} and Sec. \ref{07_02_threshold_sec} we consider the baseband and passband cases, respectively.


\subsection{Baseband pulses: \textit{A priori} and asymptotic thresholds w.r.t. the ATBW} \label{07_01_threshold_sec}

We consider in this subsection the baseband pulse in \eqref{unmodpulse_eq} and compute the \textit{a priori} and asymptotic thresholds w.r.t. the ATBW $\gamma$ in \eqref{atbw_eq} by considering a variable pulse width $T_w$ and a fixed \textit{a priori} domain $D_{\Theta}=[-2,2]$ ns.

\smallskip

In Fig. \ref{07_varBB_snr_Tw_pic}, we show the SQRTs of the CRLB $c$ in \eqref{crlb_toa_eq}, the maximum MSE $e_U$ in \eqref{mseU_eq}, and the MSEA $e_{\text{num}}$ in \eqref{enum_eq} w.r.t. $\rho$ and $T_w$. 
We can see that $e_{\text{num}}$ decreases as $T_w$ decreases for $\rho\geq16$ dB whereas it becomes approximately constant w.r.t.  $T_w$ for $\rho<16$ dB.
In fact, $c$ is achieved at $\rho=16$ dB (approximately equal to the asymptotic threshold), and it is also inversely proportional to $\beta_s^2$ which is in turn inversely proportional to $T_w^2$ as can be noticed from \eqref{crlb_toa_eq} and \eqref{envmqbw_gauss_eq}.
We deduce that the MSE can (resp. cannot) be reduced with baseband pulses by increasing the bandwidth (inversely proportional to the pulse width) if the available SNR is above (resp. below) the asymptotic threshold.

\smallskip

\begin{figure}
  \centering
  \includegraphics[width=8cm]{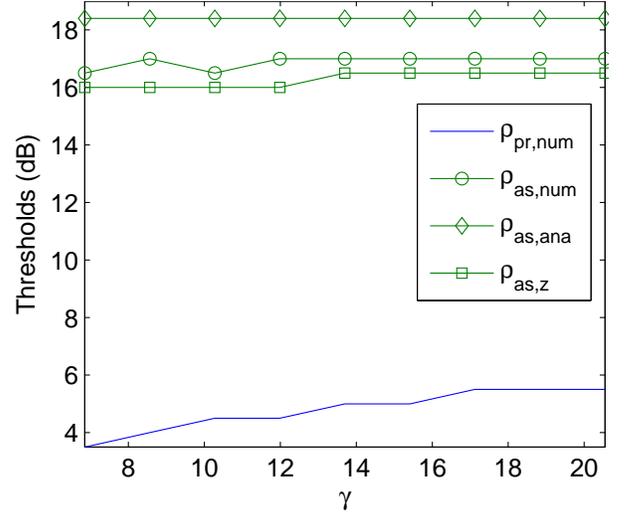}
  \caption{Baseband: \textit{A priori} and asymptotic thresholds w.r.t. the ATBW $\gamma$.}
  \label{07_thresholdBB_pic}
\end{figure}

Fig. \ref{07_thresholdBB_pic} shows the \textit{a priori} threshold $\rho_{pr,\text{num}}$ (obtained numerically from $e_{\text{num}}$), the asymptotic thresholds $\rho_{as,\text{num}}$ and $\rho_{as,z}$ (resp. obtained numerically from $e_{\text{num}}$ and the ALB $z_1$ in \eqref{z0_eq}) and the asymptotic threshold $\rho_{as,\text{ana}}$ in \eqref{thasAnal_eq} (analytic expression) w.r.t. the ATBW $\gamma$.
We can see that:
\begin{itemize}
	\item The asymptotic thresholds $\rho_{as,\text{num}}$, $\rho_{as,z}$ and $\rho_{as,\text{ana}}$ are approximately constant ($\rho_{as,\text{num}}\approx17$ dB, $\rho_{as,z}\approx16.5$ dB and $\rho_{as,\text{ana}}=18.5$ dB).
This result is already proved in Sec. \ref{thr_prop_sec}.
	\item The \textit{a priori} threshold $\rho_{pr,\text{num}}$ increases with $\gamma$; in fact, the gap between the CRLB and the maximum MSE increases with $\gamma$ while the asymptotic threshold is constant. 
\end{itemize}

\smallskip



\subsection{Passband pulses: \textit{A priori}, begin-ambiguity, end-ambiguity and asymptotic thresholds width respect to the IFBW} \label{07_02_threshold_sec}

In this subsection we consider the passband pulse in \eqref{modpulse_eq}.
We compute the \textit{a priori}, begin-ambiguity, end-ambiguity and asymptotic thresholds w.r.t. the IFBW $\lambda$ in \eqref{ifbw_eq} by considering variable pulse width $T_w$ and \textit{a priori} domain $D_{\Theta}=[-2,1.5]T_w$ and a fixed carrier $f_c=6.85$ GHz.

\smallskip

\begin{figure}
  \centering
  \includegraphics[width=8.6cm]{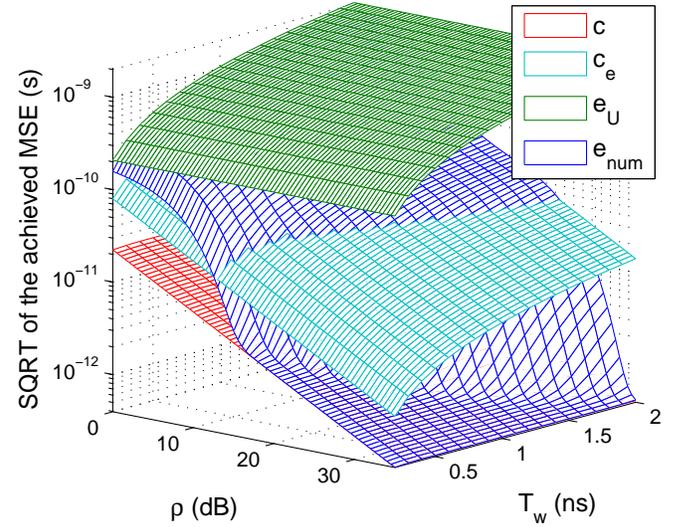}
  \caption{Passband: SQRTs of the CRLB $c$, the ECRLB $c_e$, the maximum MSE $e_U$, and the MSEA $e_{\text{num}}$ w.r.t. the SNR $\rho$ and the pulse width $T_w$.}
  \label{07_var_snr_Tw_pic}
\end{figure}

\begin{figure}
  \centering
  \includegraphics[width=8.6cm]{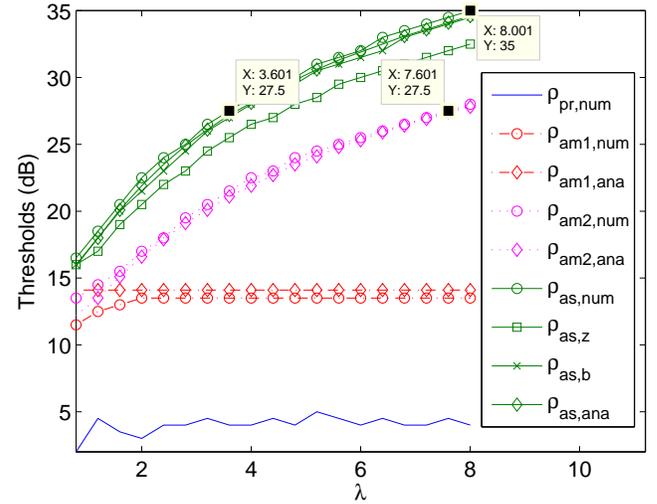}
  \caption{Passband: \textit{A priori}, begin-ambiguity, end-ambiguity, and asymptotic thresholds w.r.t. the IFBW $\lambda$.}
  \label{07_threshold_pic}
\end{figure}

In Fig. \ref{07_var_snr_Tw_pic}, we show the SQRTs of the CRLB $c$ in \eqref{crlb_toa_eq}, the ECRLB $c_e$ in \eqref{envcrlb_toa_eq}, the maximum MSE $e_U$ in \eqref{mseU_eq}, and the MSEA $e_{\text{num}}$ in \eqref{enum_eq} w.r.t. $\rho$ and $T_w$.
%
%
The ambiguity region is not observable for small $T_w$ because $e_{\text{num}}$ converges from $e_U$ to $c$ without staying long equal to $c_e$ due to the weak oscillations in the ACR; this explains why the begin-ambiguity and end-ambiguity thresholds are very close to each other for small $\lambda$ as can be seen in Fig. \ref{07_threshold_pic}.
For high $T_w$, the ambiguity region is easily observable; it has a triangular shape due to the gap between the begin-ambiguity and end-ambiguity thresholds that increases with $\lambda$ as can be seen in Fig. \ref{07_threshold_pic}.

\smallskip

In Fig. \ref{07_threshold_pic}, we show the \textit{a priori} threshold $\rho_{pr,\text{num}}$ (obtained numerically from $e_{\text{num}}$), begin-ambiguity threshold $\rho_{am1,\text{num}}$ (obtained numerically from $e_{\text{num}}$), begin-ambiguity threshold $\rho_{am1,\text{ana}}$ in \eqref{tham1Anal_eq} (analytic expression), end-ambiguity threshold $\rho_{am2,\text{num}}$ (obtained numerically from $e_{\text{num}}$), end-ambiguity threshold $\rho_{am2,\text{ana}}$ in \eqref{tham2Anal1_eq} (analytic expression), asymptotic thresholds $\rho_{as,\text{num}}$, $\rho_{as,z}$ and $\rho_{as,b}$ (resp. obtained numerically from $e_{\text{num}}$ and the ALBs $z_1$ in \eqref{z0_eq} and $b_1$ in \eqref{b0_eq}) and the asymptotic threshold $\rho_{as,\text{ana}}$ in \eqref{thasAnal1_eq} (analytic expression) w.r.t. the IFBW $\lambda$.
We can see that:
\begin{itemize}
	\item Both $\rho_{pr,\text{num}}$ and $\rho_{am1,\text{num}}$ are approximately constant. In fact, the \textit{a priori} and begin-ambiguity thresholds of a passband signal are approximately equal to the \textit{a priori} and asymptotic thresholds of its envelope (see Part 1). Furthermore, the \textit{a priori} threshold of the envelope increases with the ATBW (constant here), and its asymptotic threshold is constant (see Sec. \ref{07_01_threshold_sec}).
	\item Both $\rho_{am2,\text{num}}$ and $\rho_{as,\text{num}}$ increase with $\lambda$. In fact, the gap between the global and the local maxima of the ACR decreases as $\lambda$ increases. Therefore, a higher SNR is required to guarantee that the estimate will only fall around the global maximum.
	%
	\item The asymptotic threshold $\rho_{as,b}$ obtained from the ALB $b_1$ is very close to $\rho_{as,\text{num}}$ whereas $\rho_{as,z}$ obtained from $z_1$ is a bit far from $\rho_{as,\text{num}}$.
	\item The thresholds $\rho_{am1,\text{ana}}$, $\rho_{am2,\text{ana}}$ and $\rho_{as,\text{ana}}$ obtained from the analytic expressions are very close to $\rho_{am1,\text{num}}$, $\rho_{am2,\text{num}}$ and $\rho_{as,\text{num}}$ obtained numerically. 
	This result validates the accurateness of the analytic thresholds especially because they are obtained by considering one arbitrary envelope and by varying $f_c$ according to $\lambda$ whereas the numerical ones are obtained by varying the envelope and fixing $f_c$.
\end{itemize}

Thanks to Fig. \ref{07_threshold_pic}, we can predict the value of the achievable MSE based on the values of the available SNR and IFBW. It is approximately equal to the maximum MSE if $(\rho,\lambda)$ falls in the \textit{a priori} region (below the \textit{a priori} threshold curve), between the maximum MSE and the ECRLB if $(\rho,\lambda)$ falls in the \textit{a priori} ambiguity transition region (between the \textit{a priori} and begin-ambiguity threshold curves), approximately equal to the ECRLB if $(\rho,\lambda)$ falls in the ambiguity region (between the begin-ambiguity and end-ambiguity threshold curves), between the ECRLB and the CRLB if $(\rho,\lambda)$ falls in the ambiguity asymptotic transition region (between the end-ambiguity and asymptotic threshold curves), and approximately equal to CRLB if $(\rho,\lambda)$ falls in the asymptotic region (above the asymptotic 
threshold curve).

\smallskip  

To summarize we can say that the \textit{a priori} threshold depends on both the shape of the envelope of the ACR and the \textit{a priori} domain.
The begin-ambiguity threshold depends only on the shape of the envelope of the ACR function. 
The end-ambiguity and asymptotic thresholds only depend on the shape of the ACR, or on any set of parameters describing this shape like the shape of the envelope and the IFBW together.


\section{Signal design for minimum achievable MSE} \label{signaldesign_sec}

We have seen in Sec. \ref{threshold_sec} and Sec. \ref{thr_num_sec} that the achievable MSE depends on the available SNR and on the parameters of the transmitted signal. In this section we consider the design of the transmitted pulse spectrum w.r.t. the available SNR $\rho_0$ in order to minimize the achievable MSE.

\smallskip

We assume that the transmitted signal consists of the passband Gaussian pulse in \eqref{modpulse_eq}. 
Our goal is to find the optimal values $B_0$ and $f_{c,0}$ of the bandwidth $B$ and the carrier frequency $f_c$, respectively;
the optimal pulse width $T_{w,0}$ can be obtained from the optimal bandwidth $B_0$ using \eqref{B_gauss_eq}.

\smallskip

Regarding the constraints about the spectrum of the transmitted pulse, the two following scenarios are investigated:
\begin{enumerate}[i)]
	\item The spectrum falls in a given frequency band.
	\item The spectrum falls in a given frequency band and has a fixed bandwidth.
\end{enumerate}
The first scenario is treated in Sec. \ref{signaldesign1_sec} and the second in Sec. \ref{signaldesign2_sec}.



\subsection{Spectrum falling in a given frequency band} \label{signaldesign1_sec}

We assume in this subsection that the spectrum of the transmitted pulse falls in the frequency band $[f_l,f_h]$. This constraint can be written as:
\begin{equation} \label{constraint1_eq}
	C_1 : \left\{\begin{array}{l}
		f_c,B>0 \\
		f_c-\frac{B}{2} \geq f_l \\
		f_c+\frac{B}{2} \leq f_h. \end{array}\right.
\end{equation}
We consider the FCC UWB band\footnote{We have chosen the FCC UWB spectrum because it is possible, thanks to its ultra wide authorized band, to move the pulse spectrum around so that the IFBW be reduced and the asymptotic threshold becomes lower than or equal to the available SNR.} $[f_l,f_h]=[3.1,10.6]$ GHz \cite{fcc} in our numerical example.

\smallskip

\begin{figure}
  \centering
  \includegraphics[width=6cm]{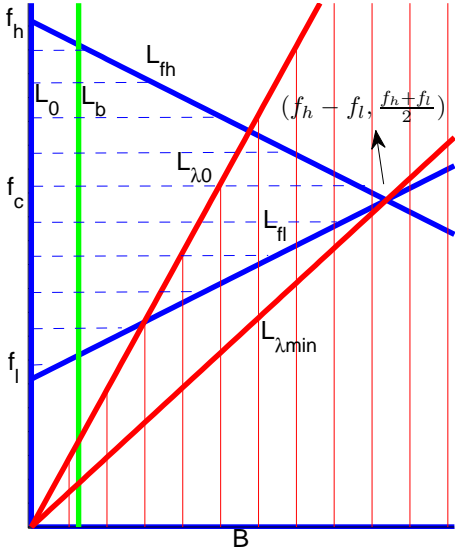}
  \caption{The feasible regions corresponding to the constraint $C_1$ in (\ref{constraint1_eq}) (region with horizontal dashed bars) and the constraint $C_2$ in (\ref{constraint2_eq}) (region with vertical solid bars).}
  \label{08_opt_prob_pic}
\end{figure}

We can write our optimization problem as:
\begin{equation}\label{opt_prob0_eq}
	(B_0,f_{c,0})=\argmin{(B,f_c)}\{e\} \text{ s.t. } \rho=\rho_0 ,\; C_1
\end{equation}
where $e$ denotes the achievable MSE. 
As depicted in Fig. \ref{08_opt_prob_pic}, the feasible region corresponding to the constraint $C_1$ in \eqref{constraint1_eq} is the triangular region (region with horizontal dashed bars) limited by the lines 
\begin{eqnarray}
	L_0 &:& B=0 \nonumber\\
	L_{f_l} &:& f_c=f_l+\frac{B}{2} \label{Lfl_eq}\\
	L_{f_h} &:& f_c=f_h-\frac{B}{2}. \label{Lfh_eq}
\end{eqnarray}
The maximum bandwidth in this feasible region is given by
\begin{equation} \label{maxB_eq}
	B_{\max} = f_h-f_l
\end{equation}
and corresponds to the intersection of the lines $L_{f_l}$ and $L_{f_h}$:
\begin{equation} \label{LflLfh_eq}
	L_{f_l}\cap L_{f_h} = \left(f_h-f_l,\frac{f_l+f_h}{2}\right).
\end{equation}
We have $B_{\max}=7.5$ GHz for the FCC UWB band.

\smallskip

For a given bandwidth $B=b$, the minimal and maximal IFBWs in the feasible region of $C_1$ are given by
\begin{eqnarray}
	\lambda_{b,\min} &=& \frac{f_l}{b}+\frac{1}{2} \label{ifbw_b_min_eq}\\
	\lambda_{b,\max} &=& \frac{f_h}{b}-\frac{1}{2} \label{ifbw_b_max_eq}
\end{eqnarray}
and correspond to the intersections of the line
\begin{equation}\label{Lb_eq}
	L_b : B=b
\end{equation}
with the lines $L_{f_l}$ and $L_{f_h}$ respectively: 
\begin{eqnarray}
	L_b\cap L_{f_l} &=& \left(b,f_l+\frac{b}{2}\right) \label{LbLfl_eq}\\
	L_b\cap L_{f_h} &=& \left(b,f_h-\frac{b}{2}\right). \label{LbLfh_eq}
\end{eqnarray}
As result, the minimal IFBW is equal to
\begin{equation}\label{min_ifbw_eq}
	\lambda_{\min} = \frac{1}{2}+\frac{f_l}{f_h-f_l}
\end{equation}
and corresponds to $L_{f_l}\cap L_{f_h}$ in \eqref{LflLfh_eq};
we have $\lambda_{\min}=0.913$ for the FCC UWB band.
The maximal IFBW is infinite and corresponds to $B=0$ GHz.

\smallskip

Let us now consider the minimization of the achievable MSE. According to the value of the available SNR $\rho_0$, three cases can be considered:
\begin{enumerate}[i)]
	\item The available SNR is lower than the begin-ambiguity threshold: $\rho_0<\rho_{am1}$; $\rho_{am1}$ is constant because it depends on the envelope shape only.
	\item The available SNR is close to the begin-ambiguity threshold: $\rho_0\approx\rho_{am1}$.
	\item The available SNR is greater than the begin-ambiguity threshold: $\rho_0>\rho_{am1}$.
\end{enumerate}

Consider the first case where $\rho_0<\rho_{am1}$. We have seen in Part 1 \cite{part1} that a passband signal and its envelope approximately achieve the same MSE below the begin-ambiguity threshold of the passband signal (approximately equal to the asymptotic threshold of the envelope). We have also seen in Sec. \ref{07_01_threshold_sec} that below the asymptotic threshold of the envelope, the achieved MSE is approximately constant and does not depend on the pulse width and the bandwidth. Therefore, nothing can be done to reduce the MSE in this case.

\smallskip

Consider the second case where $\rho_0\approx\rho_{am1}$. As the ECRLB $c_e$ in \eqref{envcrlb_toa_eq} is approximately achieved in this case, we minimize the MSE by maximizing the bandwidth $B$ (i.e. minimizing the pulse width $T_w$) so the EMQBW $\beta_e^2$ in \eqref{envcrlb_toa_eq} is maximized and $c_e$ (inversely proportional to $\beta_e^2$) is minimized. 
Therefore, the optimal solution $(B_0,f_{c,0})$ in this case and the corresponding achievable MSE $e_0$ are given by
%
%
\begin{eqnarray} \label{e0_BB_eq}
		\left\{\begin{array}{rcl}
			(B_0,f_{c,0}) &=& \left(f_h-f_l,\frac{f_l+f_h}{2}\right) \\
			e_0 &\approx& \frac{1}{\rho_0\beta_{e,0}^2} = \frac{T_{w,0}^2}{2\pi\rho_0} = \frac{2\ln10}{\pi^2B_0^2\rho_0} \end{array}\right.
\end{eqnarray}
where the expression of $e_0$ is obtained using \eqref{B_gauss_eq} and \eqref{envmqbw_gauss_eq}. Note that $f_h-f_l$ is the maximum bandwidth $B_{\max}$ in \eqref{maxB_eq}. As $\rho_{am1}\approx14$ dB as can be seen in Fig. \ref{07_threshold_pic}, we have $e_0\approx330.24$ ps$^2$ for the FCC band ($B_0=7.5$ GHz). 

\smallskip

Consider now the last case where $\rho_0>\rho_{am1}$. As we can see in Fig. \ref{07_threshold_pic}, the point $(\rho_0,\lambda)$ will fall, according to the value of the IFBW $\lambda$, in the ambiguity region, the ambiguity-asymptotic transition region, or the asymptotic region. Therefore, the achievable MSE is equal to the ECRLB $c_e$, between the ECRLB and the CRLB $c$, or equal to the CRLB.
Now, in order to find the optimal bandwidth $B_0$ and carrier $f_{c,0}$ we proceed as follows:
\begin{enumerate}
	\item We pick from Fig. \ref{07_threshold_pic} the value $\lambda_0$ of the IFBW $\lambda$ for which the available SNR $\rho_0$ belongs to the asymptotic threshold curve.
	\item In order to guarantee that the CRLB is achieved, we consider the constraint that $\lambda$ is lower than or equal to the picked $\lambda_0$. If this constraint cannot be satisfied because $\rho_0$ is lower than the minimal IFBW $\lambda_{\min}$ in \eqref{min_ifbw_eq}, then the CRLB cannot be achieved. In order to make the achievable MSE the closest possible to the CRLB, we set $\lambda$ to the minimal IFBW $\lambda_{\min}$. This constraint can be expressed as
	\begin{equation}\label{constraint2_eq}
		C_2 : \left\{\begin{array}{lll}
			\lambda = \frac{f_c}{B} \leq \lambda_0 & \text{if} & \lambda_0\geq\lambda_{\min}\\[0.1cm]
			\lambda = \frac{f_c}{B} = \lambda_{\min} & \text{if} & \lambda_0<\lambda_{\min}. \end{array}\right.
	\end{equation}
	\item Now, given that the estimator achieves the CRLB or a MSE that is the closest possible to the CRLB thanks to the previous step, we minimize the achievable MSE by minimizing the CRLB itself.
\end{enumerate}

\begin{figure*}[t]
	\begin{center}
	\subfigure[]{\includegraphics[height=5.3cm]{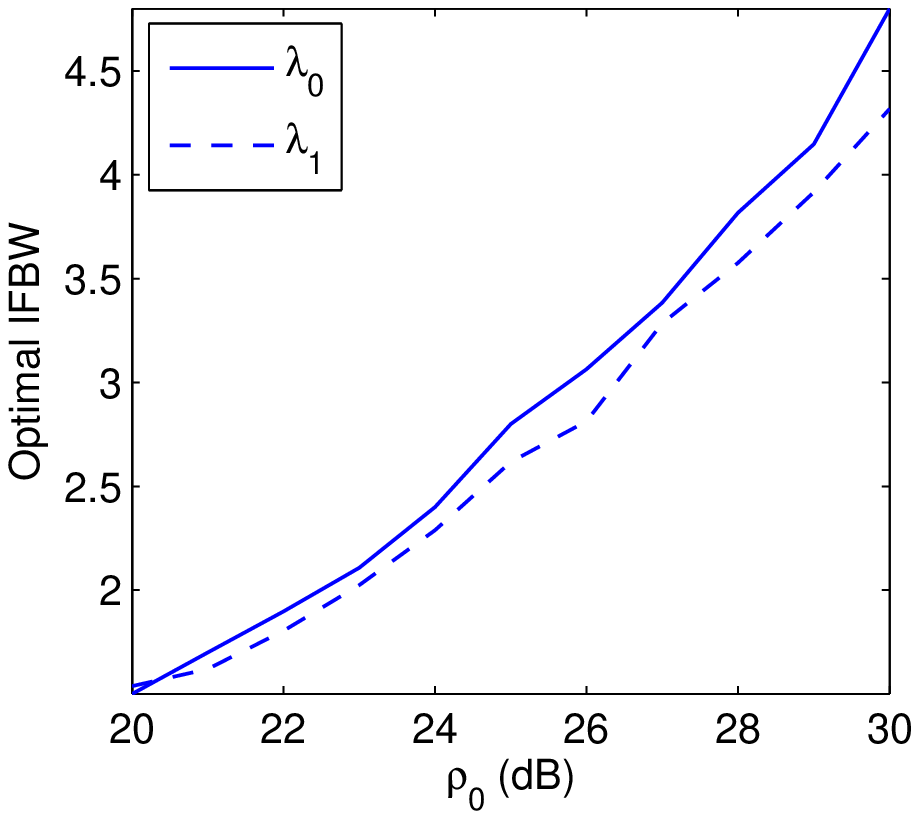} \label{08_lam0_pic}} \hspace{-0.04\textwidth}
	\quad\subfigure[]{\includegraphics[height=5.3cm]{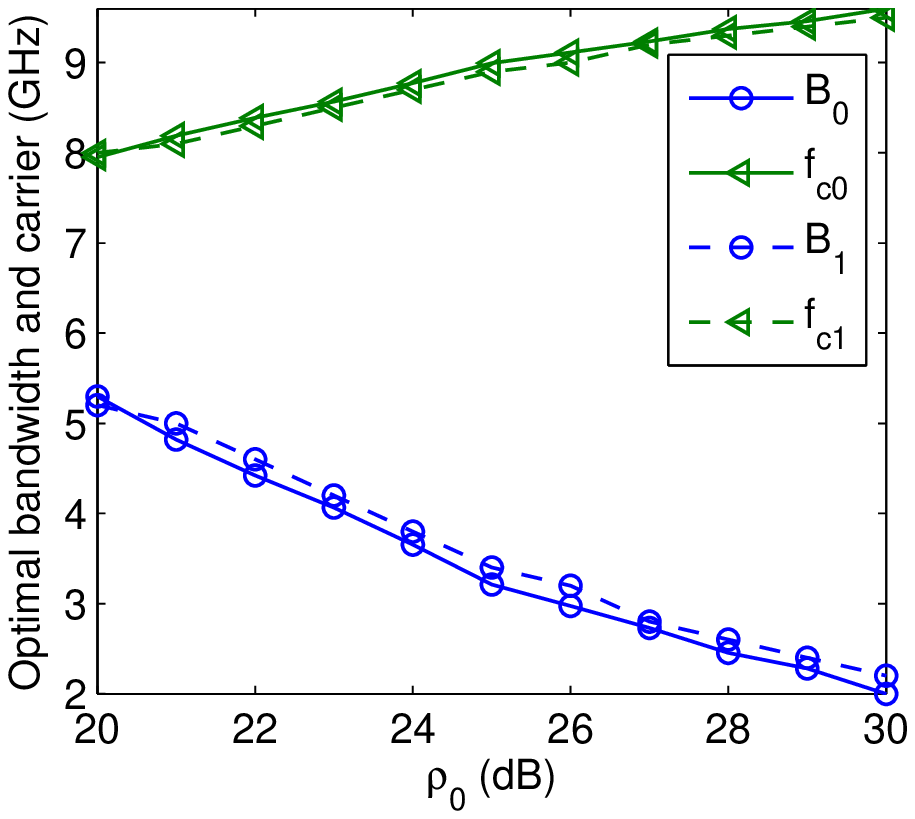} \label{08_B0_fc0_pic}} \hspace{-0.04\textwidth}
	\quad\subfigure[]{\includegraphics[height=5.3cm]{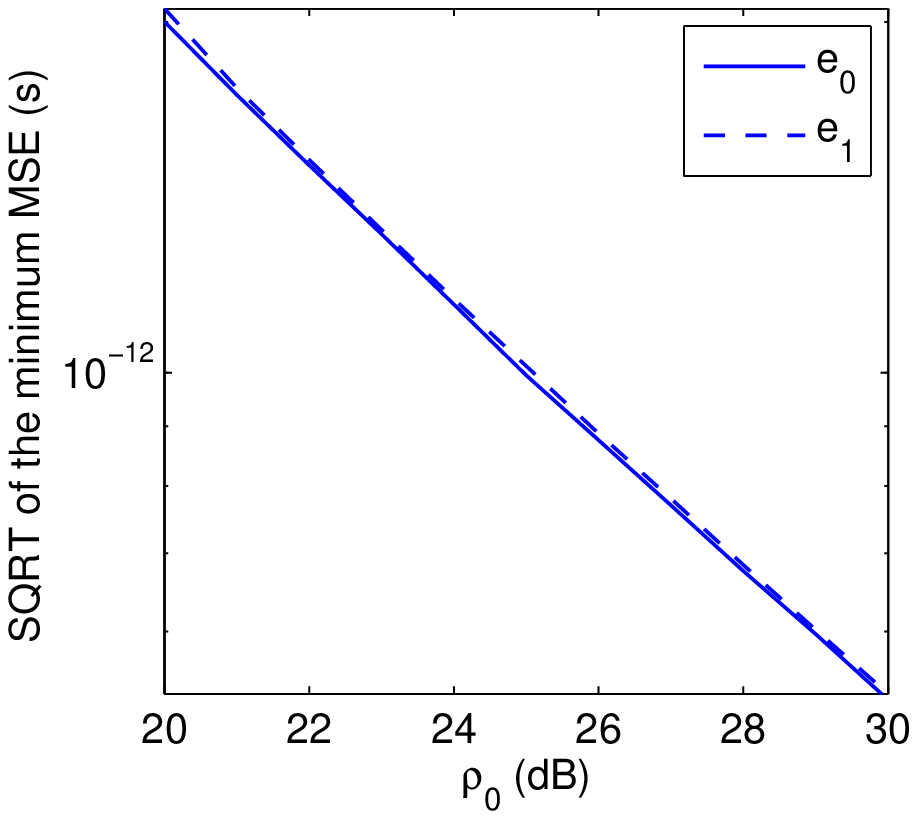} \label{08_mse0_pic}}
	\caption{(a) Suboptimal $\lambda_0$ and optimal $\lambda_1$ IFBW w.r.t. the available SNR $\rho_0$
	(b) Suboptimal $(B_0,f_{c,0})$ and optimal $(B_1,f_{c,1})$ bandwidth and carrier frequency w.r.t. $\rho_0$
	(c) SQRTs of the suboptimal $e_0$ and optimal $e_1$ MSE w.r.t. $\rho_0$.}
	\end{center}
\end{figure*}

According to the last step, we can write from \eqref{constraint1_eq} and \eqref{constraint2_eq} the minimization problem in \eqref{opt_prob0_eq} as
\begin{equation}\label{opt_prob1_eq}
	(B_0,f_{c,0})=\argmin{(B,f_c)}\{c\} \text{ s.t. } C_1 , C_2.
\end{equation}
As $c$ can be approximated from \eqref{crlb_toa_eq} and \eqref{mqbw_toa_eq} by
\begin{equation}\label{crlb8_eq}
	c = \frac{1}{\rho\beta^2_s} = \frac{1}{\rho(\beta^2_e+4\pi^2f_c^2)} \approx \frac{1}{\rho4\pi^2f_c^2}
\end{equation}
we can write the minimization problem in \eqref{opt_prob1_eq} as
\begin{equation}\label{opt_prob2_eq}
	(B_0,f_{c,0})=\argmax{(B,f_c)}\{f_c\} \text{ s.t. } C_1 , C_2.
\end{equation}
As shown in Fig. \ref{08_opt_prob_pic}, the feasible region of the constraint $C_2$ in \eqref{constraint2_eq} is the half-space below the line $L_{\lambda_0}:f_c=\lambda_0B$ (region with vertical solid bars). We have already seen that the feasible region of the constraint $C_1$ in \eqref{constraint1_eq} is the triangle limited by the lines $L_0$, $L_{f_l}$ and $L_{f_h}$. Therefore, the feasible region of $C_1$ and $C_2$ together is the triangular region limited by $L_{f_l}$, $L_{f_h}$ and $L_{\lambda_0}$ (region with both vertical and horizontal bars). 
Consequently, the solution of the maximization problem in \eqref{opt_prob2_eq} corresponds to the point of intersection $(\frac{2}{2\lambda_0+1}f_h,\frac{2\lambda_0}{2\lambda_0+1}f_h)$ of the lines $L_{f_h}$ and $L_{\lambda_0}$ as can easily be seen 
in Fig. \ref{08_opt_prob_pic}.
In the special case where $\lambda_0<\lambda_{\min}$, the feasible region of $C_2$ reduces to the line $L_{\lambda_{\min}}:f_c=\lambda_{\min}B$ so the feasible region of $C_1$ and $C_2$ reduces to the point $(f_h-f_l,\frac{f_l+f_h}{2})$ which is as result the solution of \eqref{opt_prob2_eq}.

\smallskip


Finally, the solution when the available SNR is larger than the begin-ambiguity threshold and the corresponding achievable MSE are given by:
\begin{equation}\nonumber
	\begin{array}{lll}
		\left\{\begin{array}{rcl}
			(B_0,f_{c,0}) &=& \left(f_h-f_l,\frac{f_l+f_h}{2}\right)\\
			e_0 &\in& \left]\frac{1}{4\pi^2f_{c,0}^2\rho_0},\frac{2\ln10}{\pi^2B_0^2\rho_0}\right[ \end{array}\right. &\text{if}& \lambda_0<\lambda_{\min}\\
		\left\{\begin{array}{rcl}
			(B_0,f_{c,0}) &=& \left(\frac{2f_h}{2\lambda_0+1},\frac{2\lambda_0f_h}{2\lambda_0+1}\right)\\
			e_0 &=& \frac{1}{4\pi^2f_{c,0}^2\rho_0} \end{array}\right. &\text{if}& \lambda_0\geq\lambda_{\min}
	\end{array}
\end{equation}
with $\frac{1}{4\pi^2f_{c,0}^2\rho_0}$ being the CRLB at the SNR $\rho_0$, and $\frac{2\ln10}{\pi^2B_0^2\rho_0}$ the minimum MSE in \eqref{e0_BB_eq} achieved when $\rho_0\approx\rho_{am1}$.

\smallskip

\smallskip

Let us now discuss a numerical example about the scenario considered in this subsection.
We denote by $(B_1,f_{c,1})$ the point minimizing the MSEA $e_{\text{num}}$ in the band $[f_l,f_h]=[3.1,10.6]$ GHz, $e_1$ the minimal $e_{\text{num}}$, and $\lambda_1$ the corresponding IFBW. 
To obtain $(B_1,f_{c,1})$, $e_1$ and $\lambda_1$, the available band is swept (exhaustive search) using an increment of 0.2 GHz for the bandwidth $B$ and 0.1 GHz for the carrier $f_c$.

\smallskip

In Fig. \ref{08_lam0_pic} we show $\lambda_0$ (obtained from our method) and $\lambda_1$, both w.r.t. the available SNR $\rho_0$. We can see that $\lambda_1$ is a bit smaller than $\lambda_0$. This is due to the factor $\alpha_{as}=1.1$ in the definition of the asymptotic threshold in \eqref{thas_eq}.
For $\rho_0=22$ dB, we have $\lambda_0=1.9$ and $\lambda_1=1.8$.

\smallskip

In Fig. \ref{08_B0_fc0_pic} we show $B_0$ and $f_{c,0}$ (obtained from our method), and $B_1$ and $f_{c,1}$ w.r.t. $\rho_0$. We can see that $B_0$ and $f_{c,0}$ are very close to $B_1$ and $f_{c,1}$, respectively. This result shows that our solution is very close to the optimal one. We can also see that $B_1$ (resp. $f_{c,1}$) is a bit larger (resp. lower) than $B_0$ (resp. $f_{c,0}$). In fact, $\lambda_1\lessapprox\lambda_0$ as already observed in Fig. \ref{08_lam0_pic}.
For $\rho_0=22$ dB, we have $(B_0,f_{c,0})=(4.42,8.39)$ GHz and $(B_1,f_{c,1})=(4.6,8.3)$ GHz.

\smallskip

In Fig. \ref{08_mse0_pic} we show the SQRTs of $e_0$ (minimum MSE obtained from our method) and $e_1$ w.r.t. $\rho_0$. We can see that $e_0$ and $e_1$ are very close to each other.
For $\rho_0=22$ dB, we have $e_0=2.27$ ps$^2$ and $e_1=2.32$ ps$^2$.


\subsection{Spectrum falling in a given frequency band and having a fixed bandwidth} \label{signaldesign2_sec}

We assume here that the spectrum of the transmitted pulse falls in the frequency band $[f_l,f_h]$ and has the fixed bandwidth $B=b$.
The constraint about the bandwidth can be written as:
\begin{equation} \label{constraint3_eq}
	C_3 : B=b.
\end{equation}
The feasible region corresponding to the constraints $C_1$ in \eqref{constraint1_eq} and $C_3$ in \eqref{constraint3_eq} is the segment of the line $L_b$ in \eqref{Lb_eq} limited by the lines $L_{f_l}$ in \eqref{Lfl_eq} and $L_{f_h}$ in \eqref{Lfh_eq}; 
in this feasible region, the IFBW satisfies:
\begin{equation} \nonumber
	\lambda \in [\lambda_{b,\min},\lambda_{b,\max}]
\end{equation}
where $\lambda_{b,\min}$ is given in \eqref{ifbw_b_min_eq} and $\lambda_{b,\max}$ in \eqref{ifbw_b_max_eq}.

\smallskip

To minimize the MSE, the available SNR $\rho_0$ should fall in the asymptotic region; accordingly, we write the following constraint similarly to the constraint $C_2$ in \eqref{constraint2_eq}:
\begin{equation}\label{constraint4_eq}
	C_4 : \left\{\begin{array}{lll}
		\lambda = \frac{f_c}{B} = \lambda_{b,\min} & \text{if} & \lambda_0<\lambda_{b,\min}\\[0.1cm]
		\lambda = \frac{f_c}{B} \leq \lambda_0 & \text{if} & \lambda_{b,\min}\leq\lambda_0\leq\lambda_{b,\max}\\[0.1cm]
		\lambda = \frac{f_c}{B} = \lambda_{b,\max} & \text{if} & \lambda_0>\lambda_{b,\max}. \end{array}\right.
\end{equation}
Our optimization problem can be formulated as:
\begin{equation}\label{opt_prob3_eq}
	(B_0,f_{c,0}) = \argmax{(B,f_c)}\{f_c\} \text{ s.t. } C_1 , C_3 , C_4.
\end{equation}
\\[-0.25cm]
The solution of \eqref{opt_prob3_eq} is $L_b\cap L_{f_l}$ in \eqref{LbLfl_eq} for $\lambda_0<\lambda_{b,\min}$, $L_b\cap L_{f_h}$ in \eqref{LbLfh_eq} for $\lambda_0>\lambda_{b,\max}$, and
\begin{equation}\label{LbLlam0_eq}
	L_b\cap L_{\lambda_0} = (b,\lambda_0b)
\end{equation}
for $\lambda_{b,\min}\leq\lambda_0\leq\lambda_{b,\max}$.

\smallskip

We can write the solution of our optimization problem and the corresponding achievable MSE as:
\begin{equation}\nonumber
	\begin{array}{lll}
		\left\{\begin{array}{rcl}
			(B_0,f_{c,0}) &=& \left(b,f_l+\frac{b}{2}\right) \\
			e_0 &\in& I_{e_0} \end{array}\right. &\text{if}& \lambda_0<\lambda_{b,\min}\\
		\left\{\begin{array}{rcl}
			(B_0,f_{c,0}) &=& (b,\lambda_0b) \\
			e_0 &=& \frac{1}{4\pi^2f_{c,0}^2\rho_0} \end{array}\right. &\text{if}& \lambda_{b,\min}\leq\lambda_0\leq\lambda_{b,\max}\\
		\left\{\begin{array}{rcl}
			(B_0,f_{c,0}) &=& \left(b,f_h-\frac{b}{2}\right) \\
			e_0 &=& \frac{1}{4\pi^2f_{c,0}^2\rho_0} \end{array}\right. &\text{if}& \lambda_0>\lambda_{b,\max}
	\end{array}
\end{equation}
with $I_{e_0} = \;\left]\frac{1}{4\pi^2f_{c,0}^2\rho_0},\frac{2\ln10}{\pi^2B_0^2\rho_0}\right[$.

\medskip

To apply our method, the receiver should measure the SNR and send the estimate to the transmitter, unless if the latter can estimate the SNR by itself like with mono-static radar.

\smallskip

In Sec. \ref{signaldesign1_sec} and Sec. \ref{signaldesign2_sec} we have considered two typical examples. More setups with other pulse shapes (we follow the same procedure for any carrier-modulated pulse) and with other constraints can be investigated as well. The solution of any optimization problem suffering from threshold and ambiguity effects consists in general in two steps:
\begin{enumerate}
	\item Define w.r.t. to the parameters of the considered problem the feasible region where the CRLB achieved.
	\item Minimize the CRLB by taking into account the different constraints.
\end{enumerate}
In Examples 1 and 2 below, we illustrate numerically based on the optimization problem in Sec. \ref{signaldesign2_sec} the improvement provided by each of the two steps mentioned above. 

\smallskip


\subsubsection{Example 1} 

For $\rho_0=27.5$ dB, we can see from Fig. \ref{07_threshold_pic} that $\rho_{as,\text{num}}=\rho_0$ for $\lambda=3.6$ and $\rho_{am2,\text{num}}=\rho$ for $\lambda=7.6$. So if $b=1$ GHz, then by choosing $f_c=3.6$ GHz (resp.  $7.6$ GHz) the achieved RMSE is approximately equal to $\sqrt{e_1}=\sqrt{1.1c}=2$ ps (resp. $\sqrt{e_2}=\sqrt{0.5c_e}=10\sqrt{e_1}=20$ ps). The estimation accuracy is highly improved because the CRLB is achieved instead of the ECRLB (first optimization step).

\smallskip


\subsubsection{Example 2}

For $\rho_0=35$ dB, Fig. \ref{07_threshold_pic} shows that $\rho_{as,\text{num}}=\rho_0$ for $\lambda=8$; 
so by choosing $f_c=8$ GHz (resp.  $3.6$ GHz) the achieved RMSE is approximately equal to $\sqrt{e_1}=\sqrt{1.1c_1}=0.4$ ps (resp. $\sqrt{e_2}=\sqrt{c_2}=2\sqrt{e_1}=0.8$ ps). The RMSE becomes 2 times smaller thanks to the minimization of the CRLB (second optimization step). The maximum possible improvement of the second step is $\frac{f_h-\frac{b}{2}}{\alpha_{as}(f_l+\frac{b}{2})}$ ($2.675$ for $\alpha_{as}=1.1$, $b=1$ GHz and $[f_l,f_h]=[3.1,10.6]$ GHz).


\smallskip

Let us consider a third example.

\smallskip

\subsubsection{Example 3}

Assume now that the measured SNR is 27.5 dB whereas the true one is 35 dB; then, based on the results of Example 2, the achieved RMSE will be 2 times larger. 

\smallskip

We have mentioned in Sec. \ref{intro_sec} that optimal time-bandwidth product design is considered in \cite{Van1968} based on the MIE; the mentioned work is based on the probability of non-ambiguity rather than the MSE. Therefore, the obtained solution is optimal only for sufficiently high SNRs (as supposed therein). 

\smallskip

In this section we have one main contribution.
We have considered an optimization problem subject to the threshold and ambiguity phenomena. We have proposed a very simple algorithm that minimizes the achievable MSE. 
To the best of our knowledge, this work has never been done before. 
The obtained solution is completely different from the one obtained by minimizing the CRLB (e.g, \cite{signaldesign}). When the threshold and ambiguity phenomena are not taken into account, then the optimal solution consists in filling the available spectrum with the maximum allowed PSD starting from the highest frequency.
The works in \cite{Van1968,signaldesign} correspond to the second step of our optimization method. 

\smallskip

Finally, we would like to point out that the results of Sec. \ref{signaldesign_sec} might be useful in practical UWB-based positioning systems (e.g, outdoor applications) where the multipath component resolvability, as well as the perfect multiuser interference suppression, can be insured.


\section{Conclusion}

We have employed the MIE-based MSEA to derive analytic expressions for the begin-ambiguity, end-ambiguity and asymptotic thresholds. 
The obtained thresholds are very accurate, and also can be used with various estimation problems.
We have proved that the begin-ambiguity threshold only depends on the shape of the ACR envelope, and the end-ambiguity and asymptotic thresholds only on the shape of the ACR.
Therefore, the asymptotic threshold is constant for baseband pulses with a given shape (e.g, Gaussian, cardinal sine, raised cosine with constant roll-off).
For passband pulses with given envelope shape, the begin-ambiguity threshold is constant whereas the end-ambiguity and asymptotic thresholds are functions of the IFBW.
We have exploited the information on the begin-ambiguity and asymptotic thresholds to optimize, according to the available SNR, the pulse spectrum that achieves the minimum attainable MSE. The proposed method is very simple and very accurate.




\footnotesize
\bibliographystyle{IEEEtranN}
\bibliography{IEEEabrv,all_chapters_ref_abbrev}


\end{document}